\documentclass[article]{elsarticle}

\usepackage{amssymb}
\usepackage{lineno}
\modulolinenumbers[5]

\journal{Physics Letters A}

\begin{document}

\begin{frontmatter}
\title{Consistency of the structure of Legendre transform in thermodynamics with the Kolmogorov-Nagumo average}

\author{A.M. Scarfone}
\address{Istituto dei Sistemi Complessi (ISC-CNR) c/o
Politecnico di Torino\\ Corso Duca degli Abruzzi 24, 10129 Torino, Italy.}
\ead{antoniomaria.scarfone@cnr.it}

\author{H. Matsuzoe}
\address{Department of Computer Science and Engineering, Nagoya Institute of Technology\\
Nagoya 466-8555, Japan}

\author{T. Wada}
\address{Department of Electrical and Electronic Engineering, Ibaraki University\\ Nakanarusawacho, Hitachi 316-8511, Japan}

\begin{abstract}
We show the robustness of the structure of Legendre transform in thermodynamics against the replacement of the standard linear average with the
Kolmogorov-Nagumo nonlinear average to evaluate the expectation values of the macroscopic physical observables. 
The consequence of this statement is twofold: 1)
the relationships between the expectation values and the corresponding Lagrange multipliers still hold in the present formalism; 2) the universality of the Gibbs
equation as well as other thermodynamic relations are unaffected by the structure of the average used in the theory.
\end{abstract}


\end{frontmatter}


\section{Introduction}

In the last decades, the interest toward generalized entropic forms is growing after the more and more large evidence that power-law distributions are recurrently
found in the description of the statistical proprieties of several complex systems running from the physical and biological realm to the social and economical
sciences. In order to account for such phenomenologies several entropic forms have been proposed.\\ Basically, the most of them can be grouped in two different
large classes. Often, they are obtained as continuous deformation of the well-known Shannon or Boltzmann-Gibbs (SBG) entropy, by means of suitable deformation
parameters. The first class corresponds to the trace-like entropic forms, defined as the linear average of an appropriate Hartley information function that represents the
elementary information gain
\begin{eqnarray}
  S[p]=\langle I(p)\rangle_{\rm lin} \ ,\label{trace}
\end{eqnarray}
where $\langle x\rangle_{\rm lin}=\sum_ip_i\,x_i$. In particular, for the Shannon information, the Hartley function is given by $I(p)=-\ln p$. The Sharma-Tanja-Mittal entropy \cite{STM}, defined in
\begin{eqnarray}
  S^{\rm STM}_{\alpha\gamma}[p]=\langle I_{\alpha\gamma}(p)\rangle_{\rm lin} \ ,\label{STM}
\end{eqnarray}
belongs to this class. In Eq. (\ref{STM})
\begin{eqnarray}
  I_{\alpha\gamma}(p_i)=-{p_i^\alpha-p_i^\gamma\over\alpha-\gamma} \ ,
\end{eqnarray}
where $\alpha$ and $\gamma$ are two real deformation parameters and hereinafter, $p=\{p_i\geq0;\,\sum_ip_i=1;\,(i=1,\ldots,\,W)\}$ denotes a normalized discrete distribution function.

The second class is given by the kernel-like entropic forms defined in
\begin{eqnarray}
  S[p]=K\left(\langle I(p)\rangle_{\rm lin}\right) \ ,\label{kernel}
\end{eqnarray}
that are functions of trace-like entropic forms. The Sharma-Mittal entropy \cite{SM}, defined in
\begin{eqnarray}
  S^{\rm SM}_{\alpha\gamma}[p]=K_{\alpha\gamma}\left(\langle I_\alpha(p)\rangle_{\rm lin}\right) \ ,\label{SM}
\end{eqnarray}
belongs to this class. In Eq. (\ref{SM}) 
\begin{eqnarray}
  I_\alpha(p_i)=p_i^{\alpha-1} \ ,
\end{eqnarray}
and the kernel function $K_{\alpha\gamma}(x)$, is given by
\begin{eqnarray}
K_{\alpha\gamma}(x)={x^{1-\gamma\over1-\alpha}-1\over1-\gamma} \ .
\end{eqnarray}
Both the entropy families (\ref{STM}) and (\ref{SM}) have been introduced previously in information theory and then rediscovered in statistical mechanics. Several
one-parameter deformations of the SBG entropy introduced in statistical physics belong to these families. Among them, the Kaniadakis-entropy, the Abe-entropy and the Tsallis-entropy are special
cases of the STM-entropy (see \cite{noi} and references therein), whilst the second family embodies the Tsallis-entropy, the R\'{e}nyi entropy, the escort
entropy, the Landsberg-Vedral entropy and others (see \cite{AMS} and reference therein).\\
Often, in statistical mechanics, the approach to obtain the equilibrium distribution of the system, is based on the variational principle \cite{Jaynes}. It consists in maximizing the entropic form under suitable constraints given by the values of several thermodynamic variables. Consistence with definitions (\ref{trace}) and
(\ref{kernel}) requires that constraints are introduced by means of linear average.

A different approach to look at no-trace-form entropies, often ignored in the field of the statistical mechanics, exists. It is based on the
nonlinear (o quasi-linear) average originally advanced by Kolmogorov and Nagumo \cite{KN} and then generalized by de Finetti \cite{F}. The Kolmogorov-Nagumo
(KN) average reads
\begin{eqnarray}
\langle x\rangle_{\rm KN}=f^{-1}\left(\sum_ip_i\,f(x_i)\right) \ ,\label{KN}
\end{eqnarray}
where $f(x)$ is an arbitrary strictly monotonic and continuous function, first time employed in statistics by R\'{e}nyi in the seminal work \cite{Renyi}. Today, the KN average is recurrently used in the framework of statistical mathematics and information theory \cite{Eguchi}, however, as already observed, it is often ignored in statistical mechanics. \\
In analogy with Eq. (\ref{STM}), one could define a kernel-like entropy by means of a quasi-linear trace-like form with the KN average for a suitable Hartley
information function, that is
\begin{eqnarray}
  S[p]=\langle\tilde I(p)\rangle_{\rm KN} \ .\label{SKN}
\end{eqnarray}
The two definitions (\ref{kernel}) and (\ref{SKN}) are equivalent each other for $K(x)\equiv f^{-1}(x)$ and $I(x)=f(\tilde I(x))$.

Actually, as suggested by some Authors \cite{Czachor,Bagci}, if we employ the KN average in the definition of the entropy, then the same
prescription should be used, for consistence, in the construction of the constraints used in the variational problem. This may change significantly the shape of
the distribution at equilibrium.

In order to fruitfully introduce the KN average in statistical mechanics and thermodynamics, we wonder which impact on the epistemological structure of the emerging theory it has. The structure of the Legendre transform of the theory is an important aspect to be considered.
This is a fundamental question that joins the two theories since, the Legendre transform structure binds the phenomenological aspect of thermodynamics with the microscopical aspect of statistical mechanics and, for a successful application of the KN average in statistical mechanics, such mathematical structure
should be reproduced in some way.\\
In the past, the robustness of the Legendre transform structure has been shown to hold \cite{Plastino} when the thermostatistic theory is based on trace-like entropy and the thermodynamic observables are defined by linear averages. This result has been generalized in \cite{Yamano} for the most general expression of the constraints including, in principle, the KN average. However, at the best of our knowledge, no such investigation has been done when the entropic form is of kernel-like, as in Eq. (\ref{SKN}). Our mayor effort in this work is to elucidate this problem.

The plan of the paper is as it follows. In the next section 2, we revisit the Jaynes variational principle in the framework of the KN nonlinear average, while, in section 3, we derive the structure of Legendre transform of thermodynamics based on the KN average and we obtain the corresponding Gibbs
equation and other related thermostatistic relations. A concluding summary is given in the
final section 4 whilst in Appendix A, as an example, we explicit our results in the case of the SM family.

\section{Jaynes variational principle in the framework of Kolmogorov-Nagumo average}

Let us consider a statistical system described by the entropic form (\ref{SKN}),
with a suitable Hartley function $\tilde I(x)$ that we suppose to be monotonic and normalizable but not necessarily additive. By introducing the function
$\Lambda(x)=f_1(I(x))$, we rewrite the entropy (\ref{SKN}) in
\begin{eqnarray}
  S[p]=f_1^{-1}\left(\langle\Lambda(p)\rangle_{\rm lin}\right) \ ,\label{ent}
\end{eqnarray}
where $\Lambda(x)$ is assumed to be a strictly positive, monotonic decreasing $d\Lambda(x)/dx<0$ and convex $d^2\Lambda(x)/dx^2>0$ function.

Let ${\cal O}^{(j)}$, with $j=1,\ldots,\,M$, be a set of macroscopical quantities that, with abuse of language, we call thermodynamic variables.
Consistently with definition (\ref{SKN}), we pose ${\cal O}^{(j)}=\langle o^{(j)}\rangle_{\rm KN}$ so that, they correspond to the KN averages of $M$ random variables,
where $o_{i_j}^{(j)}\,(i_j=1,\ldots,\,n_j$, with $n_j\gg M$), are the respective possible outcomes. Thus
\begin{eqnarray}
  {\cal O}^{(j)}=f_2^{-1}\left(\left\langle f_2\left(o^{(j)}\right)\right\rangle_{\rm lin}\right) \ ,\label{vin1}
\end{eqnarray}
where, for sake of exposition, we have introduced in Eqs. (\ref{ent}) and (\ref{vin1}) two different Kolmogorov functions, although in general $f_1(x)= f_2(x)\equiv f(x)$.\\
By posing $\theta_i^{(j)}=f_2\left(o_{i_j}^{(j)}\right)$, the $M$ constraints (\ref{vin1}) can be written as functions of the linear average of $\theta^{(j)}_{i_j}$, or equivalently
\begin{eqnarray}
  f_2\left({\cal O}^{(j)}\right)=\langle \theta^{(j)}\rangle_{\rm lin} \ .\label{vin}
\end{eqnarray}
Therefore, the KN average and the linear average are related each other according to
\begin{eqnarray}
  f_2\left(\langle x\rangle_{\rm KN}\right)=\langle f_2(x)\rangle_{\rm lin} \ .
\end{eqnarray}
In the same way, we can introduce the trace-like entropy
\begin{eqnarray}
 \Sigma[p]=f_1(S[p])\equiv\langle\Lambda(p)\rangle_{\rm lin} \ .\label{ass}
\end{eqnarray}
We call $\Sigma[p]$ the associated entropic form to the kernel-like entropy $S[p]$.

In order to obtain the probability distribution for the system at equilibrium, we apply the Jaynes maximum entropy principle that consists in solving the extremal problem for the maximum of the entropy $S[p]$, under the $M+1$ constraints given by the thermodynamic variables ${\cal O}^{(j)}$ and the normalization
\begin{eqnarray}
\sum_{\{i\}}p_{\{i\}}=1 \ .\label{norm}
\end{eqnarray}
In the remainder, for sake of exposition, we employ a multi-index notation with $p_{\{i\}}\equiv p_{i_1,i_2,\ldots i_M}$ and summation run over all possible combinations, that is
\begin{eqnarray}
\sum_{\{i\}}p_{\{i\}}\equiv\sum_{i_1,i_2,\ldots i_M}p_{i_1,i_2,\ldots i_M} \ .
\end{eqnarray}
In this notation ${\{i\}}=(i_1,\ldots,\,i_M)$ is an unordered $M$-tuple of integers, with entries $1\leq i_j\leq n_j$ indicating the possible outcome of $j$-th observable $o_{i_j}^{(j)}$.\\
Therefore, let $\beta_j$ and $\beta_0$ be the respective $M+1$ Lagrange multipliers, we consider the following variational problem
\begin{eqnarray}
 \delta\left[f_1^{-1}\left(\sum_{\{i\}}p_{\{i\}}\,\Lambda\big(p_{\{i\}}\big)\right)
 -\beta_0\sum_{\{i\}}p_{\{i\}}-\sum_j\beta_j f_2^{-1}\left(\sum_{\{i\}}p_{\{i\}}\,\theta^{(j)}_{i_j}\right)\right]=0 \ ,
\end{eqnarray}
that is solved for the unknown quantities $p_{\{i\}}$. The Lagrange multipliers, that ultimately turn out to be function only of the quantities
${\cal O}^{(j)}$, may be eventually obtained by using constraints (\ref{vin1}) and (\ref{norm}).\\
By performing the variation, we get
\begin{eqnarray}
F_1(S)\left(\Lambda(p_{\{i\}})+p_{\{i\}}\,{\partial\Lambda(p_{\{i\}})\over\partial p_{\{i\}}}\right)-\beta_0
-\sum_j\beta_j\,F_2\left({\cal O}^{(j)}\right)\,\theta_{i_j}^{(j)}=0 \ ,\label{maxent}
\end{eqnarray}
where the functions $F_i(x)$ are defined in
\begin{eqnarray}
  F_i(x)={\partial f_i^{-1}(y)\over\partial y}\Big|_{y=f_i(x)} \ .\label{dKN}
\end{eqnarray}
They reduce to identity $F_i(x)=1$ in the case of linear average $f_i(x)=x$.\\
By posing
\begin{eqnarray}
h(x)=\Lambda(x)+x\,{d\,\Lambda(x)\over dx} \ ,\label{h}
\end{eqnarray}
that is a strictly monotonics function for $x>0$ and then invertible, we can solve formally Eq. (\ref{maxent}) according to
\begin{eqnarray}
  p_{\{i\}}=h^{-1}\left(\tilde\beta_0+\sum_j\tilde\beta_j\,\theta_{i_j}^{(j)}\right) \ ,\label{dis}
\end{eqnarray}
where
\begin{eqnarray}
  \tilde\beta_0={\beta_0\over F_1(S)} \ ,\qquad{\rm}\qquad\tilde\beta_j=\beta_j\,{F_2\left({\cal O}^{(j)}\right)\over F_1(S)} \ .\label{tl}
\end{eqnarray}
At this step, Eq. (\ref{dis}) turns out to be a function of $2\,M+2$ quantities: the entropy $S$, the Lagrange multipliers $\beta_0$ and $\beta_j$ and the
thermodynamic variables ${\cal O}^{(j)}$.
Clearly, this is a consequence of the nonlinear flavor of the average used in the definition of the entropy and the constraints. In a theory based on the standard linear
average, $f_i(x)=x$ and therefore $F_i(x)=1$, so that $\tilde\beta\to\beta$ and the dependence from the entropy disappears.

It is worth to observe that, being $f_1(x)$ a monotonic function, the variational problem for the associate entropy $\Sigma[p]$ gives the same
distribution (\ref{dis}), although the relationships between the Lagrange multipliers $\beta_0$ and $\beta_j$, and the corresponding constraints, differ in the two
cases.
In fact, from the following variational problem
\begin{eqnarray}
 \delta\left[\sum_{\{i\}}p_{\{i\}}\,\Lambda\big(p_{\{i\}}\big)
 -\beta_0\sum_{\{i\}}p_{\{i\}}-\sum_j\beta_j f_2^{-1}\left(\sum_{\{i\}}p_{\{i\}}\,\theta^{(j)}_{i_j}\right)\right]=0 \ ,
\end{eqnarray}
it is straightforward to obtain the distribution
\begin{eqnarray}
  p_{\{i\}}=h^{-1}\left(\beta_0+\sum_j\bar\beta_j\,\theta_{i_j}^{(j)}\right) \ ,\label{dis1}
\end{eqnarray}
that is formally equal to Eq. (\ref{dis}) while $\bar\beta_j=\beta_j\,F_2({\cal O}^{(j)})$.
Finally, if we use, in this case, the linear average for constraints coherently with the trace form of entropy $\Sigma[p]$, we obtain a simplest expression for the distribution given by $p_{\{i\}}=h^{-1}(\beta_0+\sum_j\beta_j\,o_{i_j}^{(j)})$.

\section{Legendre transform structure, Gibbs equation and the reciprocity relations}

In standard thermostatistics, the Legendre structure of the theory establishes a deep connection between
thermodynamics and statistical mechanics. From one hand, it is introduced to derive potential functions
to study particular thermodynamics transformations. On the other hand, it is used to establish a relationships 
between the thermodynamic variables that, in the orthodox theory are assumed to be extensive quantities, and their conjugate variables, 
corresponding to the Lagrange multipliers, that in the orthodox theory are assumed to be intensive quantities.

A different interpretation of the Legendre transform structure arises in the framework of the geometric thermodynamics. As firstly suggested by Hermann \cite{Hermann}, thermodynamic manifold can be considered like a $2\,M+1$ dimensional space, derived from the
thermodynamic principles \cite{Arnold}. The equilibrium configuration is then represented by a $M$ dimensional sub-manifold, named Legendre manifold, where only $M$ quantities, typically identified with the thermodynamic variables ${\cal O}^{(j)}$, are assumed to be independent. They play the r\^{o}le of coordinates. In the Legendre manifold, the Gibbs equation holds. Otherwise, the remaining $M+1$ variables, given by the Lagrange multipliers $\beta_j$ and the entropy $S$, are assumed to be functions of ${\cal O}^{(j)}$. In this geometrical pictures, the Legendre transform introduces a kind of coordinate transformation in the Legendre manifold, so that one can define new thermodynamic
functions on the equilibrium manifold where the r\^{o}le of the independent quantities is, partially or completely, interchanged.

This interpretation of the Legendre transform and its related structure still holds in the framework of the KN average if the equilibrium distribution is derived from a variational principle.
To show this, let us firstly describe, in a formal way, how distribution (\ref{dis}) can be surrounded into the Legendre manifold. Apparently, this distribution
depends on $2\,M+2$ variables. However, these quantities are not all independent since $p_i$ must satisfy $M+1$ constrained relations. Among them, the
normalization condition is saturated by solving equation (\ref{norm}) for $\beta_0$. Therefore, $\beta_0=\beta_0(\beta_1,\ldots,\,\beta_n)\equiv\beta_0[\beta]$
becomes a function of the other Lagrange multipliers. In addition, the $M$ relations (\ref{vin1}) can be solved for the corresponding $\beta_j$ so that,
$\beta_j=\beta_j\left[{\cal O}^{(1)},\ldots,\,{\cal O}^{(M)};\,S\right]\equiv\beta_j\left[{\cal O};\,S\right]$. At this step, each Lagrange multiplier
depends only on the thermodynamic variables and the entropy. However, by inserting distribution (\ref{dis}) in Eq. (\ref{ent}), we can derive an implicit relation
for $S[p]$, that once solved, gives us the entropy as function of the thermodynamic variables only: $S[p]\equiv S\left[{\cal O}\right]$ and, ultimately, $\beta_j\equiv\beta_j\left[{\cal O}\right]$.

In order to derive the Legendre transform structure of the theory, we firstly evaluate the derivatives of entropy with respect to $\beta_j$
\begin{eqnarray}
\nonumber
{\partial S[p]\over\partial\beta_j}&=&\sum_{\{i\}}{\partial S[p]\over\partial p_{\{i\}}}\,{\partial p_{\{i\}}\over\partial\beta_j}\\
&=&F_1(S)\,\sum_{\{i\}}\left(\Lambda\left(p_{\{i\}}\right)+p_{\{i\}}\,{\partial\Lambda\left(p_{\{i\}}\right)
\over\partial p_{\{i\}}}\right)\,{\partial p_{\{i\}}\over\partial\beta_j} \ ,
\end{eqnarray}
and accounting for definition (\ref{h}) and Eqs. (\ref{dis}) and (\ref{tl}), we obtain
\begin{eqnarray}
\nonumber
{\partial S[p]\over\partial\beta_j}&=&F_1(S)\,\sum_{\{i\}}h(p_{\{i\}})\,{\partial p_{\{i\}}\over\partial\beta_j}\\
\nonumber
&=&\sum_{\{i\}}\left(\beta_0+\sum_k\beta_k\,F_2\left({\cal O}^{(k)}\right)\,\theta_{i_k}^{(k)}\right)\,{\partial p_{\{i\}}\over\partial\beta_j}\\
&=&\sum_k\beta_k\,F_2\left({\cal O}^{(k)}\right)\,\left(\sum_{\{i\}}\theta_{i_k}^{(k)}\,
\,{\partial p_{\{i\}}\over\partial\beta_j}\right) \ ,\label{26}
\end{eqnarray}
where we used also the relation $\sum_{\{i\}} dp_{\{i\}}=0$, which follows from condition (\ref{norm}). On the other hand, from Eq. (\ref{vin1}), rewritten in
\begin{eqnarray}
  {\cal O}^{(k)}=f_2^{-1}\left(\sum_{\{i\}}p_{\{i\}}\theta^{(k)}_{i_k}\right) \ ,
\end{eqnarray}
we have
\begin{eqnarray}
  {\partial{\cal O}^{(k)}\over\partial\beta_j}&=&F_2\left({\cal O}^{(k)}\right)\sum_{\{i\}}\theta_{i_k}^{(k)}\,{\partial p_{\{i\}}\over\partial\beta_j} \ ,
\end{eqnarray}
that, inserted in relation (\ref{26}), gives us the generalized Euler equations
\begin{eqnarray}
{\partial S\left[{\cal O}\right]\over\partial\beta_j}=\sum_k\beta_k\,{\partial{\cal O}^{(k)}\over\partial\beta_j} \ .
\end{eqnarray}
By multiplying both side by $\partial\beta_j/\partial{\cal O}^{(i)}$, and summing on the index $j$, we finally obtain
\begin{eqnarray}
  {\partial S\left[{\cal O}\right]\over\partial{\cal O}^{(i)}}=\beta_i \ ,\label{S0}
\end{eqnarray}
that establish the conjugate character between the Lagrange multipliers and the mean values.
Therefore, $\beta_j$ and ${\cal O}^{(j)}$ are related each other like the intensive versus extensive variables of thermodynamic, although they have
not longer necessarily the same intensive/extensive character.

Actually, a straightforward derivation of Eqs. (\ref{S0}) can be given by using the relation
$\sum_{\{i\}}\theta_{i_j}^{(j)}\,dp_{\{i\}}=df_2\left({\cal O}^{(j)}\right)$, as it follows from Eq. (\ref{vin}). Therefore, we have
\begin{eqnarray}
\nonumber
dS\left[{\cal O}\right]&=&F_1(S)\,\sum_{\{i\}}h(p_{\{i\}})\,dp_{\{i\}}=\sum_j\beta_j\,F_2\left({\cal O}^{(j)}\right)\left(\sum_{\{i\}}\theta_{i_j}^{(j)}\,dp_{\{i\}}\right)\\
\nonumber
&=&\sum_j\beta_j\,F_2\left({\cal O}^{(j)}\right)\,df_2\left({\cal O}^{(j)}\right)\\
&=&\sum_j\beta_j\,F_2\left({\cal O}^{(j)}\right)\,
{df_2\left({\cal O}^{(j)}\right)\over d{\cal O}^{(j)}}\,d{\cal O}^{(j)} \ ,
\end{eqnarray}
and observing that
\begin{eqnarray}
  {df_2\left({\cal O}^{(j)}\right)\over d{\cal O}^{(j)}}={1\over F_2\left({\cal O}^{(j)}\right)} \ ,
\end{eqnarray}
we obtain
\begin{eqnarray}
dS\left[{\cal O}\right]=\sum_j\beta_j\,d{\cal O}^{(j)} \ ,\label{S1}
\end{eqnarray}
that corresponds to the Gibbs equation for the present formalism. Accounting for the independence of $d{\cal O}^{(j)}$, we recover again Eq. (\ref{S0}).

By means of a partial Legendre transform we can introduce up to $2^M-2$ different functions called Massieu potentials.
They follow from any possible partition $I+J$ of the set of indices $\{1,\ldots,\,M\}$ into two disjoint subsets. Thus, function
\begin{eqnarray}
  \Phi[{\cal O}^{(I)},\,\beta_J]=S[{\cal O}]-\sum_J\beta_J\,{\cal O}^{(J)} \ ,
\end{eqnarray}
depend on the thermodynamic variables ${\cal O}^{(i)}$, with $i\in I$, and the Lagrange multipliers $\beta_j$, with $j\in J$. In fact
\begin{eqnarray}
  d\Phi[{\cal O}^{(I)},\,\beta_J]&=&dS[{\cal O}]-\sum_Jd\beta_J\,{\cal O}^{(J)}-\sum_J\beta_J\,d{\cal O}^{(J)} \ ,
\end{eqnarray}
and recalling Eq. (\ref{S1}), we obtain
\begin{eqnarray}
  d\Phi[{\cal O}^{(I)},\,\beta_J]-\sum_I\beta_I\,d{\cal O}^{(I)}+\sum_J{\cal O}^{(J)}\,d\beta_J=0 \ ,\label{psi1}
\end{eqnarray}
or equivalently
\begin{eqnarray}
  {\partial\over\partial{\cal O}^{(i)}}\Phi[{\cal O}^{(I)},\,\beta_J]=\beta_i \ ,\qquad\qquad{\partial\over\partial\beta_j}\Phi[{\cal O}^{(I)},\,\beta_J]
  =-{\cal O}^{(j)} \ ,\label{psi2}
\end{eqnarray}
that state again the duality between the Lagrange multipliers and the thermodynamic variables.

In contrast, a total Legendre transform, with $J=\{1,\ldots,\,M\}$, defines a thermodynamic potential $\psi[\beta]$, that is a function of all Lagrange multipliers $\beta_i$
\begin{eqnarray}
  \psi[\beta]=S[{\cal O}]-\sum_j\beta_j\,{\cal O}^{(j)} \ ,\label{psi3}
\end{eqnarray}
with
\begin{eqnarray}
  d\psi[\beta]=-\sum_j{\cal O}^{(j)}\,d\beta_j \ .\label{GDg}
\end{eqnarray}
The role of dependent and independent variables is exchanged so that, ${\cal O}^{(j)}\equiv{\cal O}^{(j)}[\beta]$ are now the dependent variables, functions of the Lagrange multipliers $\beta_i$ that turn out to be the independent variables. Therefore, Eq. (\ref{GDg}) implies
\begin{eqnarray}
  {\partial\psi[\beta]\over\partial\beta_i}=-{\cal O}^{(i)} \ .\label{psi4}
\end{eqnarray}

Many thermodynamic relations hold also in the present formalism.
For instance, from Eqs. (\ref{S0}) and (\ref{psi4}), it is straightforward to obtain
\begin{eqnarray}
  {\partial{\cal O}^{(j)}\over\partial\beta_i}={\partial{\cal O}^{(i)}\over\partial\beta_j} \ ,
\end{eqnarray}
that, in the standard thermodynamic, correspond to the Maxwell relations. These equations imply the following relationships between the potentials $S[{\cal O}]$ and $\psi[\beta]$:
\begin{eqnarray}
  {\partial^2S[{\cal O}]\over\partial{\cal O}^{(i)}\,\partial{\cal O}^{(j)}}=\left({\partial^2\psi[\beta]\over\partial\beta_i\,\partial\beta_j}\right)^{-1} \ .\label{Hessian}
\end{eqnarray}
In the framework of information geometry the two Hessian matrices appearing in Eq.(\ref{Hessian}) define, respectively, the metric tensor
\begin{eqnarray}
g^{ij}={\partial^2 S[{\cal O}]\over\partial{\cal O}^{(i)}\,\partial{\cal O}^{(j)}} \ ,
\end{eqnarray}
and its inverse
\begin{eqnarray}
g_{ij}={\partial^2 \psi[\beta]\over\partial\beta_i\,\beta_j} \ ,
\end{eqnarray}
of the corresponding statistical manifold \cite{noi1} that, according to Eq. (\ref{Hessian}), is equivalents to
\begin{eqnarray}
g^{ik}\,g_{kj}=\delta^i_j \ .
\end{eqnarray}

Proceeding further, we consider again the entropy (\ref{ent}) in the probability space and looking at the following relation
\begin{eqnarray}
  {\partial S[p]\over \partial p_{\{i\}}}=\ell_{\{i\}}[p] \ ,\label{dual}
\end{eqnarray}
where
\begin{eqnarray}
  \ell_{\{i\}}[p]=F_1(S)\,h\big(p_{\{i\}}\big) \ .
\end{eqnarray}
According to Eq. (\ref{dual}), $\big(p_{\{i\}},\,\ell_{\{i\}}\big)$ forms a set of
canonically conjugate variables in the statistical manifold, like as like, the set $({\cal O}^{(j)},\,\beta_j)$ does in the Legendre manifold. This suggest us to
introduce a Legendre transform of the entropy in the statistical manifold \cite{noi1}, to obtain a new functional ${\cal S}[\ell]$
\begin{eqnarray}
{\cal S}[\ell]=S[p]-\sum_{\{i\}}p_{\{i\}}\,\ell_{\{i\}}[p] \ .
\end{eqnarray}
However, taking into account that $\ell_{\{i\}}$ actually depend on $p_{\{i\}}$, we can pull-back the functional ${\cal S}[\ell]$ onto the probability space and define the new quantity ${\cal I}^\ast[p]$, where ${\cal I}^\ast[p]\equiv{\cal I}\big[\beta[p]\big]$, as
\begin{eqnarray}
\nonumber
  {\cal I}^\ast[p]&\equiv&{\cal S}[(\ell[p])]\\
  \nonumber
  &=&S[p]-F_1(S[p])\,\langle h(p_{\{i\}})\rangle_{\rm lin}\\
  &=&S[p]-\beta_0-\sum_j\beta_j\,F_2\left({\cal O}^{(j)}\right)\,\langle\theta^{(j)}\rangle_{\rm lin} \ .
\end{eqnarray}
By using Eq. (\ref{psi3}), we introduce the pull-back of the $\psi$-potential $\psi^\ast[p]\equiv\psi\big[\beta[p]\big]$, according to
\begin{eqnarray}
 \psi^\ast[p]={\cal I}^\ast[p]+\beta_0+\sum_j\beta_j\,F_2\left({\cal O}^{(j)}\right)\,{\cal D}\left(\theta^{(j)}\right) \ ,\label{psi5}
\end{eqnarray}
where function ${\cal D}(x)$ is defined in
\begin{eqnarray}
{\cal D}(x)=\langle x\rangle_{\rm lin}-
{\langle x\rangle_{\rm KN}\over F_2\left(\langle x\rangle_{\rm KN}\right)} \ ,\label{D}
\end{eqnarray}
and measures the discrepancy between the KN average and the standard linear average. In fact, it is easy to see as ${\cal D}(x)=0$ when $f_2(x)=x$, that is the natural setup for trace-form entropies. In this case, the expression of the $\psi$-potential simplifies in
\begin{eqnarray}
  \psi^\ast[p]={\cal I}^\ast[p]+\beta_0 \ .\label{psi8}
\end{eqnarray}
The same relation holds also for the associate entropy $\Sigma[p]$ introduced in Eq. (\ref{ass}), where the ${\cal I}$-function is given by
\begin{eqnarray}
  {\cal I}^\ast[p]=\Sigma[p]-\sum_{\{i\}}p_{\{i\}}\,h\big(p_{\{i\}}\big) \ ,
\end{eqnarray}
since, in this case, $\ell\big(p_{\{i\}}\big)\equiv h\big(p_{\{i\}}\big)$.\\
It could be of a certain interest to observe as the function ${\cal D}(x)$ can be introduced in
\begin{eqnarray}
  {\cal D}(x)=f_2(x)-x\,{df_2(x)\over dx} \ ,
\end{eqnarray}
that is reminiscent of a Legendre transform of the Kolmogorov function $f_2(x)$.

\section{Final summary}

In this letter, we proved in the formulation of a thermostatistic theory the robustness of the structure of Legendre transform against the replacement of the standard linear average with the Kolmogorov-Nagumo nonlinear average. This result can be summarized up as it follows
\begin{eqnarray}
\nonumber
\begin{array}{lcl}
S[{\cal O}]-\psi[\beta]+\sum_j\beta_j\,{\cal O}^{(j)}=0& &\\
& &\\
dS[{\cal O}]-\sum_j\beta_j\,d{\cal O}^{(j)}=0&\Rightarrow\qquad&{\partial S[{\cal O}]\over\partial{\cal O}^{(j)}}=\beta_j \\
& &\\
d\psi[\beta]-\sum_j{\cal O}^{(j)}\,d\beta_j=0&\Rightarrow\qquad&{\partial\psi[\beta]\over\partial\beta_j}=-{\cal O}^{(j)} \\
& &\\
\left.\begin{array}{l}
{\partial^2S[{\cal O}]\over\partial{\cal O}^{(i)}\,\partial{\cal O}^{(j)}}={\partial\beta_i\over\partial{\cal O}^{(j)}}={\partial\beta_j\over\partial{\cal O}^{(i)}}\\
\\
{\partial^2\psi[\beta]\over\partial\beta_i\,\partial\beta_j}={\partial{\cal O}^{(j)}\over\partial\beta_i}={\partial{\cal O}^{(i)}\over\partial\beta_j}
\end{array}
\right\}
&\Rightarrow\qquad&
{\partial^2S[{\cal O}]\over\partial{\cal O}^{(i)}\,\partial{\cal O}^{(j)}}=\left({\partial^2\psi[\beta]\over\partial\beta_i\,\partial\beta_j}\right)^{-1}
\end{array}
\end{eqnarray}
It is shown, one more time, that the structure of Legendre transform of thermodynamics is strictly related to the maximal entropy principle used to derive the equilibrium distribution, confirming and extending in this way the discussion advanced in \cite{Plastino,Yamano}. An important consequence of our result concerns the Gibbs equation of thermodynamics that, as showed, still holds also in the present formalism.

\appendix
\section{}

In this appendix we specify the results obtained in the paper by considering, as an explicit example, the kernel-like entropic form belonging to the SM family (\ref{SM}). It can be derived from the KN average according to
\begin{eqnarray}
S^{\rm SM}_{\alpha\gamma}[p]=\varphi^{-1}_{\alpha\gamma}\left(\sum_ip_i\,\varphi_{\alpha\gamma}
\left(I_\gamma(p_i)\right)\right)\equiv
{\Big(\sum_{\{i\}}p^\alpha_{\{i\}}\Big)^{1-\gamma\over1-\alpha}-1\over1-\gamma} \ .\label{SM1}
\end{eqnarray}
where the Kolmogorov function $f_1(x)=f_2(x)=\varphi_{\alpha\gamma}(x)$ and the elementary information gain $\tilde I(p)\equiv I_\gamma(p)$ are given, respectively, by
\begin{eqnarray}
  \varphi_{\alpha\gamma}(x)=\ln_\alpha\left(\exp_\gamma(x)\right) \ ,
\end{eqnarray}
and
\begin{eqnarray}
 I_\gamma(p_i)=\ln_\gamma(1/p_i) \ ,
\end{eqnarray}
where $\ln_\alpha(x)$ and $\exp_\alpha(x)$, with $\ln_\alpha(\exp_\alpha(x))=\exp_\alpha(\ln_\alpha(x))=x$, are the deformed logarithm and exponential
\cite{Tsallis}, defined in
\begin{eqnarray}
  &&\ln_\alpha(x)={x^{1-\alpha}-1\over1-\alpha} \ ,\\
  &&\exp_\alpha(x)=(1+(1-\alpha)\,x)^{1\over1-\alpha} \ .
\end{eqnarray}
The above two definitions reduce to standard logarithm and exponential in the $\alpha\to1$ limit, i.e $\ln_1(x)=\ln(x)$ and $\exp_1(x)=\exp(x)$.

For simplicity, we consider a canonical system with $M=1$, where the only constraint, in addition to the normalization of $p_i$, is given by the energy average
$U=\langle\epsilon\rangle_{\rm KN}$.\\ From the variational problem
\begin{eqnarray}
\nonumber
  &&\delta\left[{1\over1-\gamma}\left(1+\sum_ip_i^\alpha-\sum_ip_i\right)^{1-\gamma\over1-\alpha}\right.\\
  &-&\left.\beta_0\sum_ip_i- {\beta_1\over1-\gamma}\left(1+(1-\alpha)\sum_ip_i\,\theta_i\right)^{1-\gamma\over1-\alpha}
\right]=0 \ ,
\end{eqnarray}
with $\theta_i=\ln_\alpha\left(\exp_\gamma(\epsilon_i)\right)$, we obtain
\begin{eqnarray}
  {1\over1-\alpha}\left(\sum_i p_i^\alpha\right)^{\alpha-\gamma\over1-\alpha}\left(\alpha\,p_i^{\alpha-1}-1\right)-\beta_0
  -\beta_1\,\left(\exp_\alpha\left(\langle\theta\rangle_{\rm lin}\right)\right)^{\alpha-\gamma}\,\theta_i=0 \ ,\label{e1}
\end{eqnarray}
that corresponds to Eq. (\ref{maxent}), with $h(x)=(\alpha\,x^{\alpha-1}-1)/(1-\alpha)$ and $F_1(x)=F_2(x)\equiv\big(\exp_\gamma(x)\big)^{\alpha-\gamma}$. By
solving Eq. (\ref{e1}) for $p_i$, we get the distribution in the form
\begin{eqnarray}
  p_i=c_\alpha\,\Big[1-(\alpha-1)\,(\tilde\beta_0+\tilde\beta_1\,\theta_i)\Big]^{1\over\alpha-1} \ ,\label{SMdist}
\end{eqnarray}
where $c_\alpha=\alpha^{1\over1-\alpha}$ is a constant.\\
Unless $\gamma=1$, distribution (\ref{SMdist}) has an asymptotic power-law behavior for $\theta_i\gg1$ (i.e. $\epsilon_i\gg1$). Otherwise,
for $\gamma=1$, corresponding to the R\'{e}nyi entropy
\begin{eqnarray}
  S_\alpha^{\rm R}[p]\equiv S_{\alpha\gamma=1}^{\rm SM}[p]={1\over1-\alpha}\,\ln\sum_i p_i^\alpha \ ,
\end{eqnarray}
the equilibrium distribution has an exponential tail, like the Gibbs one ($\alpha=\gamma=1$). This fact, already observed in \cite{Bagci}, is a consequence of the KN average used in the definition of the energy average. Differently, if we use standard linear average, $f_2(x)=x$, the equilibrium distribution (\ref{SMdist}) has always an asymptotic power-law tail whenever $\alpha\not=1$ and in particular, for the R\'{e}nyi entropy, it becomes
\begin{eqnarray}
  p_i=c_\alpha\,\left[1-(\alpha-1)\,\big(\exp\left(S_\alpha\right)\big)^{1-\alpha}
  \left(\beta_0+\beta_1\,\epsilon_i\right)\right]^{1\over\alpha-1} \ ,
\end{eqnarray}
with a power-law tail \cite{Bashkirov}.\\
The associated entropy to $S_{\alpha\gamma}^{\rm SM}$:
\begin{eqnarray}
  \Sigma_\alpha[p]=\varphi_{\alpha\gamma}\left(S_{\alpha\gamma}^{\rm SM}[p]\right) \ ,
\end{eqnarray}
coincides with the Tsallis entropy
\begin{eqnarray}
  \Sigma_\alpha[p]={\sum_ip_i^\alpha-1\over1-\alpha} \ .
\end{eqnarray}
This transmutation relationship is already known in the R\'{e}nyi entropy formalism and actually holds for the whole SM family.
As a consequence, both Tsallis and SM entropies share the same equilibrium distribution given in Eq. (\ref{SMdist}) but with an appropriate redefinition of its arguments
$(\tilde\beta_0+\tilde\beta_1\,\theta_i)\to(\beta_0+\beta_1\,\epsilon_i)$.

Starting from Eq. (\ref{SM1}) we can derive the expression of the $\cal I$-function
\begin{eqnarray}
  {\cal I}^{\ast\,\rm SM}_{\alpha\gamma}[p]={1\over1-\alpha}\,\left[\left({1-\alpha\over1-\gamma}-\alpha\right)
  \,z_\alpha[p]^{1-\gamma}+z_\alpha[p]^{\alpha-\gamma}\right]-{1\over1-\gamma} \ ,
\end{eqnarray}
with ${\cal I}^{\ast\,\rm SM}_{\alpha\gamma}\to{\cal I}^{\ast\,\rm SM}_{00}\equiv{\cal I}^{\ast\,\rm BG}=1$, where $z_\alpha[p]$ is the partition function of the system \cite{AMS}, given by $z_\alpha[p]^{1-\alpha}=\sum_{\{i\}}p_{\{i\}}^\alpha$ and the $\psi$-potential follows directly from Eq. (\ref{psi5}).\\ In addition, from Eq. (\ref{psi3}), we can obtain the fundamental thermodynamic equation in the entropy representation \cite{Callen}, that links the entropy to the thermodynamic variables
\begin{eqnarray}
  \Big(\exp_\gamma\left(S^{\rm SM}_{\alpha\gamma}\right)\Big)^{\alpha-\gamma}\Big(\alpha\,\varphi_{\alpha\gamma}
  \left(S^{\rm SM}_{\alpha\gamma}\right)-1\Big)=\beta_0+\sum_j\beta_j\,\overline{{\cal O}^{(j)}} \ ,\label{fundSM}
\end{eqnarray}
where, for sake of notation, we defined $\overline{{\cal O}^{(j)}}=F_2({\cal O}^{(j)})\,f_2({\cal O}^{(j)})$.\\
In the $(\alpha,\gamma)\to(1,1)$ limit, Eq. (\ref{fundSM}) reduces to the fundamental equation of the Boltzmann-Gibbs theory: $S^{\rm BG}=1+\beta_0+\sum_j\beta_j\,{\cal O}^{(j)}$.\\
In particular, for the R\'{e}nyi entropy ($\gamma=1$), the fundamental equation (\ref{fundSM}) becomes
\begin{eqnarray}
S_\alpha={1\over\alpha-1}\,\ln\Big(\alpha+(\alpha-1)\,\Big(\beta_0+\sum_j\beta_j\overline{{\cal O}^{(j)}}\Big)\Big) \ ,
\end{eqnarray}
while the $\cal I$-function is given by
\begin{eqnarray}
  {\cal I}_\alpha=\log z_\alpha-{z_\alpha^{\alpha-1}-\alpha\over\alpha-1} \ .
\end{eqnarray}
In contrast, for the associated entropic form, the $\cal I$-function is related to the entropy by an affine relation ${\cal I}^\ast_\alpha=1+(1-\alpha)\,\Sigma_\alpha$, where $\Sigma_\alpha=\varphi_{\alpha\gamma}\left(S_{\alpha\gamma}^{\rm SM}\right)$ and the fundamental equation becomes
\begin{eqnarray}
\alpha\,\Sigma_\alpha=1+\beta_0+\sum_j\beta_j\,{\cal O}^{(j)} \ .
\end{eqnarray}
A simple expression for ${\cal I}^\ast[p]$ is obtained also when $\Sigma[p]$ belongs to STM family given in Eq. (\ref{STM}). In this case,
posing $\Sigma\equiv S_{\alpha\gamma}^{\rm STM}$, the fundamental equation becomes \cite{noi2}
\begin{eqnarray}
\lambda \,S^{\rm STM}_{\alpha\gamma}\left[p^{\rm scaled}\right]=\beta_0+\sum_j\,\beta_j\,{\cal O}^{(j)} \ ,\label{struc}
\end{eqnarray}
where the scaled distribution is defined in $p^{\rm scaled}=\{p_{\{i\}}/\delta\}$ and the constants $\lambda$ and $\delta$ are given by
$\lambda^{\beta-\alpha}=(1+\alpha)^{1+\beta}/(1+\beta)^{1+\alpha}$ and $\delta^{\beta-\alpha}=(1+\alpha)/(1+\beta)$, respectively.
The scaled entropy fulfills the relation
\begin{eqnarray}
  \lambda \,S^{\rm STM}_{\alpha\gamma}\left[p^{\rm scaled}\right]=S^{\rm STM}_{\alpha\gamma}[p]-{\cal I}^{\rm STM}_{\alpha\gamma}[p] \ ,
\end{eqnarray}
where
\begin{eqnarray}
  {\cal I}^{\ast\,\rm STM}_{\alpha\gamma}[p]=\sum_{\{i\}}{p_{\{i\}}^{1+\alpha}+p_{\{i\}}^{1+\gamma}\over2}\quad
  \stackrel{(\alpha,\gamma)\to(0,0)}{\rightarrow}\quad1 \ ,
\end{eqnarray}
so that, Eq. (\ref{struc}) can be written in \cite{noi3}
\begin{eqnarray}
S^{\rm STM}_{\alpha\gamma}={\cal I}^{\ast\,\rm STM}_{\alpha\gamma}+\beta_0+\sum_j\,\beta_j\,{\cal O}^{(j)} \ .
\end{eqnarray}
\section*{Acknowledgments}

AMS is indebted with the College of Engineering, Department of Electrical and Electronic Engineering of the Ibaraki University for the warm hospitality he found during his stay. This work was supported by Italian
CNR via program STM 2015.

\vfill\eject
\end{document}